\documentclass[pra,prl,twocolumn,showpacs,floatfix,10pt,twoside]{revtex4-1}
\usepackage{graphicx}% Include figure files
\usepackage{dcolumn}% Align table columns on decimal point
\usepackage{bm}% bold math
\usepackage{subfigure}
\usepackage{float}
\usepackage{multirow}
\usepackage{booktabs}
\usepackage{amsmath}
\usepackage{latexsym,amssymb}
\usepackage[mathscr]{eucal}
\usepackage[switch*]{lineno}
\usepackage[bookmarks=true,
   colorlinks=true,
   linkcolor=blue,
   urlcolor=blue,
   citecolor=blue,
   bookmarks=true,
   hyperindex=true
]{hyperref}

\begin{document}
\allowdisplaybreaks[4]
%\preprint{APS/123-QED}

\title{Tunneling radiation of fermions from the non-stationary Kerr black hole}

\author{Qun-Chao Ding\textsuperscript{1,2,}}
\author{Zhong-Wen Feng\textsuperscript{1,2,}}
\author{Shu-Zheng Yang\textsuperscript{1,2}}
\altaffiliation{Email: szyangcwnu@126.com}
\vskip 0.5cm
\affiliation{1 Physics and Space Science College, China West Normal University, Nanchong, 637009, China\\
2 Department of Astronomy, China West Normal University, Nanchong 637009, China}

\date{\today}% It is always \today, today,
             %  but any date may be explicitly specified

\begin{abstract}
In this paper, the tunneling radiation of fermions with spin $1/2$  and $3/2$ from the non-stationary Kerr black hole are investigated. First, according to the Dirac equation for spin $1/2$ fermions and Rarita-Schwinger equation for spin $3/2$ fermions, we  construct the corresponding gamma matrixes and the derive Hamilton-Jacobi equation  for spin $1/2$ and $3/2$ fermions. Then, the tunneling behavior of fermions on the event horizon is studied. Finally, we obtain the thermodynamic properties of the non-stationary Kerr black hole.
The result shows that the tunneling  radiation rate, surface gravity and temperature are all related to ${\dot r_H}$ and ${r'_H}$.
\end{abstract}
%\pacs{04.60.Bc, 04.80.Cc, 03.75.Dg}% PACS, the Physics and Astronomy
                             % Classif\/ication Scheme.
%Use showkeys class option if keyword
                              %display desired
\maketitle
% \linenumbers
\section{Introduction}
\label{Int}
Based on  Einstein's general relativity, a lot of works have been done for studying the black hole. Those results show that general relativity is an elegant theory. Thus, the enthusiasm of investigating general relativity is aroused as hot topics. In 1974 Hawking discovered the thermodynamic of black hole \cite{ch1,ch2,ch3}. Considering the quantum ef\/fects, Hawking proved that the black holes have the thermal radiation. Hawking's thermal radiation theroy pointed out that the virtual particles inside a black hole via the quantum tunneling ef\/fect reach the event horizon and materialize real particles \cite{ch4,ch5,ch6,ch7,ch8,ch9,ch10}. In 2000, Parikh and Wilczek have put forward the tunneling method and investigated the thermal radiation from black holes by using quantum mechanics \cite{ch11}. In recent years, a lot of signif\/icant studies have been done for investigating tunneling radiation of black hole. In Refs.~ \cite{ch12,ch13,ch14,ch15,ch16,ch17}, Zhang and Zhao have studied the tunneling radiation and provided a explanation for information loss of a black hole. Later,  Lin and Yang \emph{et al.} have further researched on tunneling radiation from non-stationary black holes\cite{ch18,ch19,ch20,ch21,ch22,ch23,ch24}. In Refs.~\cite{ch25,ch26}, by taking into account the change of background of black holes, the the tunneling radiation needs to be modif\/ied. Thus, according to the Klein-Gordon equation in curved spacetimes, researchers in Refs.~\cite{ch27,ch28,ch29,ch30,ch31,ch32,ch33,ch34,ch35}  study the tunneling radiation and Hawking temperature of static, steady and dynamic black holes. Those results show that the tunneling radiation and Hawking temperature have significant correction.  As we know, the dynamics equation of spin $1/2 $ and $3/2 $ fermions in the curved spacetimes are depended on some specif\/ic matrices, which leads to complex calculations. For overcome this situation, one can use the Hamilton-Jacobi equation (HJE) to describe the dynamic features of particles in curved spacetimes. This method simplify the process of studying particles dynamic characteristics. Especially the tunneling radiation characteristics of fermions with non-zero spin in curved spacetimes of dynamic black holes.

Despite there are many researchers have been made many achievements in particles in radiation and Hawking temperature, we still need to modif\/ied it. In actually, black holes in the universe are dynamic, and it is meaningful to study the quantum tunneling radiation of fermions that spin of non-zero in the dynamic curved spacetime. In this paper, we derive the HJE from Dirac equation and Rarita-Schwinger equation of spin $1/2$ and $3/2$ fermions. Then using the advanced Eddington coordinate to describe the tunneling radiation and Hawking temperature of axially symmetric dynamic Kerr black hole.

 The rest of the paper is organized as follows.  In Sec.~\ref{sec2}, we derived the HJE from the dynamic equation of fermions.  In Sec.~\ref{sec3}, we investigated the tunneling radiation and Hawking temperature in a non-stationary symmetric cured spacetime. The last section is is devoted the discussion and conclusion.

\section{The dynamic equation of spin $1/2$ and $3/2$ fermions in a non-stationary Kerr black hole}
\label{sec2}
In the advanced Eddington coordinate, the line element of the non-stationary Kerr black hole can be written as follows \cite{ch4}:
\begin{align}
\label{eq1}
d{s^2} & =  - \left( {1 - \frac{{2Mr}}{{{\rho ^2}}}} \right)d{\upsilon^2} + 2d\upsilon dr - 2\frac{{2Mra{{\sin }^2}\theta }}{{{\rho ^2}}}d\upsilon d\varphi
 \nonumber \\
&- 2a{\sin ^2}\theta drd\varphi  + {\rho ^2}d{\theta ^2} + \left[ {\left( {{r^2} + {a^2}} \right)+ \frac{{2Mr{a^2}{{\sin }^2}\theta }}{{{\rho ^2}}}} \right]
 \nonumber \\
&\cdot {\sin ^2}\theta d{\varphi ^2},
\end{align}
where ${\rho ^2}  = {r^2} + {a^2}{\cos ^2}\theta$, $\upsilon$ is the standard advanced time. Therefore, the mass of  non-stationary Kerr black hole is time  dependent, namely, $M = M\left( \upsilon  \right)$. However, the  specif\/ic angular momentum is a  constant \cite{ch36}.  According to Eq.~(\ref{eq1}), the covariant metric tensor is given by
\begin{align}
\label{eq3}
{g_{\mu \nu }} = \left( {\begin{array}{*{20}{c}}
{{g_{00}}}&{{g_{01}}}&0&{{g_{03}}}\\
{{g_{10}}}&0&0&{{g_{13}}}\\
0&0&{{g_{22}}}&0\\
{{g_{30}}}&{{g_{31}}}&0&{{g_{33}}}
\end{array}} \right),
\end{align}
and the corresponding determinant and the inverse tensors metric are
\begin{align}
\label{eq4}
g =  - {\rho ^4}{\sin ^2}\theta ,
\end{align}
\begin{align}
\label{eq5}
{g^{\mu \nu }} = {\left( { - 1} \right)^{\mu  + \nu }}\frac{{{\Delta ^{\mu \nu }}}}{g} = \left( {\begin{array}{*{20}{c}}
{{g^{00}}}&{{g^{01}}}&0&{{g^{03}}}\\
{{g^{10}}}&{{g^{11}}}&0&{{g^{13}}}\\
0&0&{{g^{22}}}&0\\
{{g^{30}}}&{{g^{31}}}&0&{{g^{33}}}
\end{array}} \right),
\end{align}
where ${\Delta ^{\mu \nu }}$ is minors, the non-zero components of the inverse metric tensor are shown as
\begin{align}
\label{eq6}
{g^{00}} &=\frac{{{a^2}{{\sin }^2}\theta }}{{{\rho ^2}}},{g^{01}} = \frac{{{r^2} + {a^2}}}{{{\rho ^2}}},
\nonumber \\
{g^{03}}& = \frac{a}{{{\rho ^2}}},{g^{11}} = \frac{{{r^2} + {a^2} - 2Mr}}{{{\rho ^2}}},
\nonumber \\
{g^{13}}&  = \frac{a}{{{\rho ^2}}},{g^{22}} = \frac{1}{{{\rho ^2}}},{g^{33}} = \frac{1}{{{\rho ^2}{a^2}{{\sin }^2}\theta }}.
\end{align}
According to the the null-hyper surface condition ${g^{\mu \nu }}\left( {{\partial _{{x^\mu }}}F} \right)\left( {{\partial _{{x^\upsilon }}}F} \right) = 0$, where $F$ is the  hyper-surface, the position of event horizon is located at ${g^{00}}\dot r_H^2 - 2{g^{01}}{\dot r_H} + {g^{11}} + {g^{22}}{r'_H}^2 = 0$,  with ${\dot r_H} = {\partial _\nu }{r_H}$ and ${r'_H} = {\partial _\theta }{r_H}$.

Since we study the tunneling behavior of spin $1/2$ and $3/2$ fermions from the black hole, it is necessary brief\/ly  review  the Dirac equation and Rarita-Schwinger equation. It is well known that the kinematic porpoise of spin $1/2$ fermions  on the event horizon of black holes is describes by Dirac equation
\begin{align}
\label{eq7}
{\gamma ^\mu }{\mathcal{D}_\mu }\Psi  + \frac{m}{\hbar }\Psi  = 0,
\end{align}
where
\begin{align}
\label{eq8}
{\mathcal{D}_\mu } = {\partial _\mu } + \frac{i}{2}\Gamma _\mu ^{\alpha \beta }{\Pi _{\alpha \beta }},
\end{align}
\begin{align}
\label{eq9}
{\Pi _{\alpha \beta }} = \frac{i}{4}\left[ {{\gamma ^\alpha },{\gamma ^\beta }} \right].
\end{align}
Meanwhile, for studying the kinematic porpoise of spin $3/2$ fermion on the event of black holes, one needs to employ the Rarita-Schwinger equation, which can be expressed as follow:
\begin{align}
\label{eq10}
{\gamma ^\mu }{\mathcal{D}_\mu }{\Psi_\nu} + \frac{m}{\hbar }{\Psi_\nu} = 0,
\end{align}
with the supplementary condition
\begin{align}
\label{eq11}
{\gamma ^\mu }{\Psi _\nu } = 0.
\end{align}
Now, considering the spacetimes of non-stationary Kerr black hole, the gamma matrix should satisfy the relation
\begin{align}
\label{eq12}
\left\{ {{\gamma ^\mu },{\gamma ^\nu }} \right\} = 2{g^{\mu \nu}}{I}.
\end{align}
Now, combining Eq.~(\ref{eq10}) and Eq.~(\ref{eq12}), the gamma matrix can be constructed as follows:
\begin{align}
\label{eq13}
{\gamma ^\upsilon } = & {\left( {\frac{{{a^2}{{\sin }^2}\theta }}{{{\rho ^2}}}} \right)^{1/2}}\left( {\begin{array}{*{20}{c}}
{I}&0\\
0&{I}
\end{array}} \right),
\nonumber \\
{\gamma ^r}  =  & {\left[ {\frac{{{r^2} + {a^2} - 2Mr}}{{{\rho ^2}}} - \frac{{{{\left( {{r^2} + {a^2}} \right)}^2}}}{{{\rho ^2}{a^2}{{\sin }^2}\theta }}} \right]^{1/2}}
\nonumber \\
   &\cdot \left( {\begin{array}{*{20}{c}}
0&{{\sigma ^1}}\\
{{\sigma ^1}}&0
\end{array}} \right)+ \frac{{{r^2} + {a^2}}}{{\rho a\sin \theta }}\left( {\begin{array}{*{20}{c}}
{I}&0\\
0&{ - {I}}
\end{array}} \right),
\nonumber \\
{\gamma ^\theta }  = & \frac{1}{\rho }\left( {\begin{array}{*{20}{c}}
0&{{\sigma ^2}}\\
{{\sigma ^2}}&0
\end{array}} \right),
\nonumber \\
{\gamma ^\varphi } = & \frac{{{a^2}\sin \theta }}{{\rho \left( {{r^2} + {a^2}} \right)}}\left( {\begin{array}{*{20}{c}}
I&0\\
0&{ - I}
\end{array}} \right)
\nonumber \\
&  + \sqrt {\frac{1}{{{\rho ^2}{a^2}{{\sin }^2}\theta }} - \frac{{{a^4}{{\sin }^2}\theta }}{{{\rho ^2}{{\left( {{r^2} + {a^2}} \right)}^2}}}} \left( {\begin{array}{*{20}{c}}
0&{{\sigma ^3}}\\
{{\sigma ^3}}&0
\end{array}} \right),
\end{align}
where ${\sigma ^1}$, ${\sigma ^2}$ and ${\sigma ^3}$ are the Pauli matrix, and the  in the above equation represents the unit matrix. By using the gamma matrix in non-stationary Kerr black hole spacetimes, we will derive the HJE for fermions with spin $1/2$ and $3/2$ and study the fermions radiation from the non-stationary Kerr black hole.

\section{The tunneling radiation from non-stationary Kerr black hole of spin $1/2$ and $3/2$ fermions}
\label{sec3}
Due to the Dirac equation and the WKB approximation, the wave function of spin $1/2$ fermion can be expressed as
\begin{equation}
\label{eq14}
\Psi  = \xi {\rm{exp}} \left({\frac{i}{\hbar }S}\right),
\end{equation}
where $S$ is the classical action of fermions, the coef\/f\/icient term can be expressed as
\begin{align}
\label{eq15}
\xi  = \left( {\begin{array}{*{20}{c}}
{A}\\
{B}
\end{array}} \right),
\end{align}
where $A$ and $B$ are $1\times 2$ matrices. Substituting Eq.~(\ref{eq14}) into Eq.~(\ref{eq7}) and neglecting the higher order terms of $\hbar$, one  yields
\begin{align}
\label{eq16}
\left( {\begin{array}{*{20}{c}}
\alpha &\beta \\
\beta &\eta
\end{array}} \right)\left( {\begin{array}{*{20}{c}}
A\\
B
\end{array}} \right) = 0
\end{align}
The component of $\alpha $, $\beta $ and $\eta $ are shown in Eq.~(\ref{eq17}) are
\begin{align}
\label{eq17}
\alpha   = &\frac{{a\sin \theta }}{\rho }\frac{{\partial S}}{{\partial \upsilon }}{I} + \frac{{{r^2} + {a^2}}}{{\rho a\sin \theta }}\frac{{\partial S}}{{\partial r}}{I} + \frac{1}{{\rho \sin \theta }}\frac{{\partial S}}{{\partial \varphi }}{I} + im{I},
\nonumber \\
\eta   = & \frac{{a\sin \theta }}{\rho }\frac{{\partial S}}{{\partial \upsilon }}{I} + \frac{{{r^2} + {a^2}}}{{\rho a\sin \theta }}\frac{{\partial S}}{{\partial r}}{ I} + \frac{1}{{\rho \sin \theta }}\frac{{\partial S}}{{\partial \varphi }}{I} - im{ I},
\nonumber \\
\beta  = & \left[ {\frac{{{r^2} + {a^2} - 2Mr}}{{{\rho ^2}}}\frac{{\partial S}}{{\partial r}} + \frac{1}{{\rho \left( {{r^2} + {a^2} - 2Mr} \right)}}\frac{{\partial S}}{{\partial \varphi }}} \right]{\sigma ^1}
\nonumber \\
&+ \left( {\frac{1}{\rho }\frac{{\partial S}}{{\partial \theta }} + \frac{1}{{\rho a\sin \theta }}\frac{{\partial S}}{{\partial \varphi }}} \right){\sigma ^2}{\rm{ + }}\Xi,
\nonumber \\
\Xi  = & \left\{ {\frac{1}{{{\rho ^2}{{\sin }^2}\theta }} - \Delta  - \frac{1}{{{\rho ^2}{a^2}{{\sin }^2}\theta }}} \right\}\frac{{\partial S}}{{\partial \varphi }}{\sigma ^3},
 \nonumber \\
\Delta   = & \frac{{{{\left[ {{a^2}{{\sin }^2}\theta  - a\left( {{r^2} + {a^2}} \right)} \right]}^2}}}{{{\rho ^4}\left[ {{a^2}{{\sin }^2}\theta \left( {{r^2} + {a^2} - 2Mr} \right) - {{\left( {{r^2} + {a^2}} \right)}^2}} \right]}}.
 \end{align}
In order to get non-trivial solution, it requires the determination of Eq.~(\ref{eq16}) equal to zero, hence, one gets
\begin{align}
\label{eq18}
\det \left( {\alpha \eta  - \beta \beta } \right) = 0,
\end{align}
Combining the Eq.~(\ref{eq17}) with Eq.~(\ref{eq18}), which leads
\begin{align}
\label{eq19}
&\frac{{{a^2}{{\sin }^2}\theta }}{{{\rho ^2}}}{\left( {\frac{{\partial S}}{{\partial \upsilon }}} \right)^2} + 2\frac{{{r^2} + {a^2}}}{{{\rho ^2}}}\left( {\frac{{\partial S}}{{\partial r}}} \right)\left( {\frac{{\partial S}}{{\partial \upsilon }}} \right)
\nonumber \\
&+ 2\frac{a}{{{\rho ^2}}}\frac{{\partial S}}{{\partial \upsilon }}\frac{{\partial S}}{{\partial \varphi }} + \frac{{{r^2} + {a^2} - 2Mr}}{{{\rho ^2}}}{\left( {\frac{{\partial S}}{{\partial r}}} \right)^2}
 + 2\frac{a}{{{\rho ^2}}}\frac{{\partial S}}{{\partial r}}\frac{{\partial S}}{{\partial \varphi }}
\nonumber \\
&+ \frac{1}{{{\rho ^2}}}{\left( {\frac{{\partial S}}{{\partial \theta }}} \right)^2} + \frac{1}{{{\rho ^2}{a^2}{{\sin }^2}\theta }}{\left( {\frac{{\partial S}}{{\partial \varphi }}} \right)^2} + {m^2} = 0.
\end{align}
It is observably that Eq.~(\ref{eq19}) is the HJE, one can get the same result when putting the tensors into the general expressions of HJE, that is ${g^{\mu \upsilon }}\left( {{{\partial S} \mathord{\left/ {\vphantom {{\partial S} {\partial {x^\mu }}}} \right. \kern-\nulldelimiterspace} {\partial {x^\mu }}}} \right)\left( {{{\partial S} \mathord{\left/ {\vphantom {{\partial S} {\partial {x^\upsilon }}}} \right. \kern-\nulldelimiterspace} {\partial {x^\upsilon }}}} \right) + {m^2} = 0$. Notably, Eq.~(\ref{eq19}) is derived from the Dirac equation, hence, it only suitable for describing kinematic porpoise of the spin $1/2$ fermions from event horizon of the non-stationary Kerr black hole.

Next, with the same way, one can derive the HJE from Rarita-Schwinger equation. First of all, it is necessary def\/ine wave function of Eq.~(\ref{eq10}) as follows
\begin{align}
\label{eq20}
{\Psi _\nu } = \left( {\begin{array}{*{20}{c}}
{{A_\nu }}\\
{{B_\nu }}
\end{array}} \right)\exp \left( {\frac{i}{\hbar }S} \right),
\end{align}
In the above expression, we denote the matrices by ${{A}_\nu } = {\left( {\begin{array}{*{20}{c}}{{a_\nu }}&{{c_\nu }}\end{array}} \right)^{TM}},{{B}_\nu } = {\left( {\begin{array}{*{20}{c}}{{b_\nu }}&{{d_\nu }} \end{array}} \right)^{TM}}$, and ${a_\nu }$,${b_\nu }$,${c_\nu }$,${d_\nu }$ are corresponding matrix. In the semi-classical approximation, we can get
\begin{align}
\label{eq21}
\left( {\begin{array}{*{20}{c}}
\alpha &\beta \\
\beta &\eta
\end{array}} \right)\left( {\begin{array}{*{20}{c}}
{{{A}_\nu }}\\
{{{B}_\nu }}
\end{array}} \right) = 0,
\end{align}
where $\alpha $, $\beta$ and $\eta $ are shown as Eq.~(\ref{eq17}). In order to get non-trivial solution, it requires the determination of Eq.~(\ref{eq22}) equal to zero, that is
\begin{align}
\label{eq22}
\det \left( {\alpha \eta  - \beta \beta } \right) = 0.
\end{align}
The Eq.~(\ref{eq22}) leads to the following results
\begin{align}
\label{eq23}
&\frac{{{a^2}{{\sin }^2}\theta }}{{{\rho ^2}}}{\left( {\frac{{\partial S}}{{\partial \upsilon }}} \right)^2} + 2\frac{{{r^2} + {a^2}}}{{{\rho ^2}}}\left( {\frac{{\partial S}}{{\partial r}}} \right)\left( {\frac{{\partial S}}{{\partial \upsilon }}} \right) + 2\frac{a}{{{\rho ^2}}}\frac{{\partial S}}{{\partial \upsilon }}\frac{{\partial S}}{{\partial \varphi }}
\nonumber \\
&+ \frac{{{r^2} + {a^2} - 2Mr}}{{{\rho ^2}}}{\left( {\frac{{\partial S}}{{\partial r}}} \right)^2}
+ 2\frac{a}{{{\rho ^2}}}\frac{{\partial S}}{{\partial r}}\frac{{\partial S}}{{\partial \varphi }} + \frac{1}{{{\rho ^2}}}{\left( {\frac{{\partial S}}{{\partial \theta }}} \right)^2}
\nonumber \\
&+ \frac{1}{{{\rho ^2}{a^2}{{\sin }^2}\theta }}{\left( {\frac{{\partial S}}{{\partial \varphi }}} \right)^2} + {m^2} = 0,
\end{align}
which has the same expression as Eq.~(\ref{eq19}). However, the HJE from Eq.~(\ref{eq23}) is only for spin $3/2$ fermions.
Now, we can use the HJE to study the tunneling behavior of spin $1/2$ and spin $3/2$ fermions from non-stationary Kerr black hole since the their kinematic porpoise can be describe by HJE, which has the same expression. As we know, the evert horizon of non-stationary Kerr black hole varies with time $\upsilon $. Therefore, for calculate the tunneling rate of fermions on the event horizon, one needs to use the general tortoise coordinate transformation \cite{ch37}:
\begin{align}
\label{eq24}
&{r_*} = r + \frac{1}{{2\kappa }}\ln \left[ {r - {r_H}\left( {\upsilon ,\theta } \right)} \right],
\nonumber \\
&{\upsilon _*} = \upsilon  - {\upsilon _0},
\nonumber \\
&{\theta _*} = \theta  - {\theta _0},
\end{align}
where $\upsilon_0 $ and $\theta_0 $ are arbitrary constants characterizes the initial state of the hole, respectively. $\kappa$ denotes an adjustable parameter. The above-mention equations lead to
\begin{align}
\label{eq25}
&\frac{\partial }{{\partial r}} = \frac{{2\kappa (r - {r_H}) + 1}}{{2\kappa (r - {r_H})}}\frac{\partial }{{\partial {r_*}}},
\nonumber \\
&\frac{\partial }{{\partial \upsilon }} = \frac{\partial }{{\partial {\upsilon _*}}} - \frac{{\dot r_H }}{{2\kappa (r - {r_H})}}\frac{\partial }{{\partial {r_*}}},
\nonumber \\
&\frac{\partial }{{\partial \theta }} = \frac{\partial }{{\partial {\theta _*}}} - \frac{{{r'_H}}}{{2\kappa (r - {r_H})}}\frac{\partial }{{\partial {r_*}}},
\end{align}
where ${\dot r_H} = {\partial _\upsilon }{r_H}$ ,${r'_H} = {\partial _\theta }{r_H}$. Substituting Eq.~(\ref{eq25}) into line element Eq.~(\ref{eq19}) or Eq.~(\ref{eq23}), The HJE becomes
\begin{align}
\label{eq26}
&{a^2}{\sin ^2}\theta {\left( {\frac{{\partial S}}{{\partial {\theta _*}}}} \right)^2} + {a^2}{\sin ^2}\theta {\left[ {\frac{{{{\dot r}_H}}}{{2\kappa (r - {r_H})}}\frac{{\partial S}}{{\partial {r_*}}}} \right]^2}
\nonumber \\
 -& \frac{{2{a^2}{{\sin }^2}\theta }}{{2\kappa (r - {r_H})}}{{\dot r}_H}\left( {\frac{{\partial S}}{{\partial {\theta _*}}}\frac{{\partial S}}{{\partial {r_*}}}} \right) + {\left( {\frac{{\partial S}}{{\partial \theta }}} \right)^2}
 \nonumber \\
 -& \Delta {\left[ {\frac{{2\kappa (r - {r_H}) + 1}}{{2\kappa (r - {r_H})}}\frac{{\partial S}}{{\partial {r_*}}}} \right]^2}n + {\left( {\frac{{\partial S}}{{\partial {\theta _*}}}} \right)^2}
  \nonumber \\
 +& {\left[ {\frac{{{{r'}_H}}}{{2\kappa (r - {r_H})}}\frac{{\partial S}}{{\partial {r_*}}}} \right]^2} - \frac{{{{r'}_H}}}{{\kappa (r - {r_H})}}\frac{{\partial S}}{{\partial {\theta _*}}}\frac{{\partial S}}{{\partial {r_*}}}
 \nonumber \\
+ & 2\left( {{r^2} + {a^2}} \right)\left[ {\frac{{\partial S}}{{\partial {\upsilon _*}}} - \frac{{{{\dot r}_H}}}{{2\kappa (r - {r_H})}}\frac{{\partial {\rm{S}}}}{{\partial {r_*}}}} \right]\left[ {\frac{{2\kappa (r - {r_H}) + 1}}{{2\kappa (r - {r_H})}}\frac{{\partial S}}{{\partial {r_*}}}} \right]
 \nonumber \\
 + &\frac{1}{{{{\sin }^2}\theta }}{\left( {\frac{{\partial S}}{{\partial \varphi }}} \right)^2}2a\frac{{\partial S}}{{\partial \varphi }}\left[ {\frac{{2\kappa (r - {r_H}) + 1}}{{2\kappa (r - {r_H})}}\frac{{\partial S}}{{\partial {r_*}}}} \right]
 \nonumber \\
 + &{m^2}\left( {{r^2} + {a^2}{{\cos }^2}\theta } \right) = 0.
\end{align}
Simplifying the Eq.~(\ref{eq26}), one yields
\begin{align}
\label{eq27}
\frac{\mathcal{B}}{\mathcal{C}}{\left( {\frac{{\partial S}}{{\partial {r_*}}}} \right)^2} - 2\frac{{\partial S}}{{\partial {\nu _*}}}\frac{{\partial S}}{{\partial {r_*}}} - 2\frac{\mathcal{D}}{\mathcal{C}}\frac{{\partial S}}{{\partial {r_*}}} + 2\kappa \left( {r - {r_H}} \right)\frac{\mathcal{E}}{\mathcal{C}} = 0,
\end{align}
with
\begin{align}
\label{eq28}
\mathcal{B} =& \frac{{\left( {{a^2}{{\sin }^2}\theta } \right)\dot r_H^2 - 2\left( {{r^2} + {a^2}} \right){{\dot r}_H} + \left( {{r^2}{\rm{ + }}{a^2} - 2Mr} \right) + {r'_H}^2}}{{2\kappa \left( {r - {r_H}} \right)}},
\nonumber \\
\mathcal{C} = &{a^2}{\sin ^2}\theta {\dot r_H} - \left( {{r^2} + {a^2}} \right),
\nonumber \\
\mathcal{D} = & - {P_\theta }{r'_H} - a{\dot r_H}j,
\nonumber \\
\mathcal{E} =& {a^2}{\sin ^2}\theta {\left( {\frac{{\partial S}}{{\partial {\nu _*}}}} \right)^2} + {P_\theta } + \frac{{{j^2}}}{{{{\sin }^2}\theta }}- 2a{P_\theta }j
 \nonumber \\
&+ \left( {{r^2} + {a^2}{{\cos }^2}\theta } \right){m^2},
\end{align}
where ${P_\theta } = {\partial _{{\theta _*}}}S$, $j = {\partial _\phi }S$. According to Eq.~(\ref{eq27}), one obtains an inf\/inite limit of $0/0$ type on the event horizon. Therefore, it is necessary to use the L'Hopital law here, which leads to
\begin{equation}
\label{eq29}
\mathop {\lim }\limits_{\scriptstyle{\rm{r}} \to {r_H}\hfill\atop
{\scriptstyle\nu  \to {\nu _0}\hfill\atop
\scriptstyle\theta  \to {\theta _0}\hfill}} \frac{{{\mathcal{B}}}}{{{\mathcal{C}}}} = 1.
\end{equation}
By solving the Eq.~(\ref{eq29}) the $\kappa $ is given by
\begin{align}
\label{eq30}
\kappa  = \frac{{ - 2{r_H}{{\dot r}_H} + {r_H} - M}}{{\left( {1{\rm{ - }}2{{\dot r}_H}} \right)\left( {r_H^2 + {a^2}{\rm{ - }}{a^2}\dot r_H^2{{\sin }^2}{\theta _0}} \right){\rm{ + }}2{{r'}_H}^2}}.
\end{align}
Actually, $\kappa $ is the surface gravity of non-stationary Kerr black hole. On the event horizon, Eq.~(\ref{eq27}) can be rewritten as follows:
\begin{align}
\label{eq31}
{\left( {\frac{{\partial S}}{{\partial {r_*}}}} \right)^2} + 2\left( {\omega  - {\omega _0}} \right)\frac{{\partial S}}{{\partial {r_*}}} = 0,
\end{align}
where $\omega $ is the energy of the tunneling particles, and the expression of ${\omega _0}$ is denoted as
\begin{align}
\label{eq32}
{\omega _0} = \mathop {\lim }\limits_{r \to {r_H}} \frac{\mathcal{D}}{\mathcal{C}} = \frac{{ - {P_\theta }{{r'}_H} - a{{\dot r}_H}j}}{{{a^2}{{\sin }^2}\theta {{\dot r}_H} - \left( {r_H^2 + {a^2}} \right)}}.
\end{align}
Due to the solution of Eq.~(\ref{eq31}), one gets
\begin{align}
\label{eq33}
\frac{{\partial S}}{{\partial r}}& = \left[ {1 + \frac{1}{{2\kappa \left( {r - {r_H}} \right)}}} \right]\frac{{\partial S}}{{\partial {r_*}}}
 \nonumber \\
&= \frac{{\left[ {2\kappa \left( {r - {r_H}} \right) + 1} \right]\left[ {\left( {\omega  - {\omega _0}} \right) \pm \left( {\omega  - {\omega _0}} \right)} \right]}}{{2\kappa \left( {r - {r_H}} \right)}}.
\end{align}
After integral the Eq.~(\ref{eq33}), the result is
\begin{align}
\label{eq34}
S &= \int {\frac{{\left[ {2\kappa \left( {r - {r_H}} \right) + 1} \right]\left[ {\left( {\omega  + {\omega _0}} \right) \pm \left( {\omega  + {\omega _0}} \right)} \right]}}{{2\kappa \left( {r - {r_H}} \right)}}dr}  \nonumber \\
&= \frac{{i\pi }}{{2\kappa }}\left[ {\left( {\omega  - {\omega _0}} \right) \pm \left( {\omega  - {\omega _0}} \right)} \right],
\end{align}
where ${\rm{ + }}\left(  -  \right)$ mean the outgoing (incoming) solution. Taking into account both of outgoing and incoming solution, the total tunneling rate of fermions is
\begin{align}
\label{35}
\Gamma  &= {{{\Gamma _{{\rm{emission}}}}} \mathord{\left/
 {\vphantom {{{\Gamma _{{\rm{emission}}}}} {{\Gamma _{{\rm{absorption}}}}}}} \right.
 \kern-\nulldelimiterspace} {{\Gamma _{{\rm{absorption}}}}}} = \exp \left[ { - \frac{{2\pi }}{\kappa }\left( {\omega  - {\omega _0}} \right)} \right]
\nonumber \\
 & =  \exp ( - \frac{{\omega  - {\omega _0}}}{T}).
\end{align}
Obviously, the expression of $\kappa$ in Eq.~(\ref{eq30}) is the surface gravity of non-stationary Kerr black hole on the event horizon. So the Hawking temperature of non-stationary Kerr black hole is
\begin{align}
\label{36}
{T_H} & = \frac{\kappa }{{2\pi }}
\nonumber \\
&= \frac{{\left( {1 - {{\dot r}_H}} \right){r_H} - M}}{{\left( {1 - 2{{\dot r}_H}} \right)\left( {r_H^2 + {a^2} - {a^2}\dot r_H^2{{\sin }^2}{\theta _0}} \right) + 2{{r'}_H}^2}}.
\end{align}
Now, we have derived the HJE in the non-stationary Kerr black hole space-time from dynamic equation of spin $1/2$ and spin $3/2$ fermions. Then discussed the quantum tunneling and Hawking temperature. The Hawking temperature is related to ${r_H}$ and ${r'_H}$ . It should be noted that for the spin of $3/2$ fermion, we should start with the matrix Eq.~(\ref{eq9}) and in the same way to study tunneling rate and Hawking temperature.

\section{Conclusion}
\label{Dis4}
In this paper, we obtain the HJE by describing the Dirac equation of spin $1/2$ fermion and Rarita-Schwinger equation of spin $3/2$ fermion in the non-stationary curved Kerr black hole spacetime. The conclusion shows that the HJE is a basic condition for the establishment of field equation in a curved spacetimes. Based on the HJE, we have investigated the Hawking temperature and fermions tunneling radiation of axially symmetric  dynamic Kerr black hole. The result shows that the tunneling radiation, temperature and surface gravity are all related to ${r_H}$ and ${r'_H}$. For different spacetimes, there should be different gamma matrices. Therefore, this paper effectively investigated the particles tunneling radiation and Hawking temperature under the axially symmetric dynamic Kerr black hole's curved spacetime background. We can get when $a = 0,{r'_H} = 0$ the non-stationary Kerr black hole return to the case of Vaidya black hole that also proved the method that we studied is correctness.

\section*{Acknowledgements}
This work is  supported in part by the National Natural Science Foundation of China (Grant Nos. 11847048 and 11573022) and the Fundamental Research Funds of China West Normal University (Grant Nos. 17E093 and 17YC518).

% \linenumbers
\end{document}